\documentclass[lineno]{jfm}
\usepackage{verbatim}
\usepackage{graphicx}
\usepackage{newtxtext}
\usepackage{newtxmath}
\usepackage{natbib}
\usepackage{hyperref}
\hypersetup{
    colorlinks = true,
    urlcolor   = blue,
    citecolor  = black,
}

\newcommand{\RomanNumeralCaps}[1]
\linenumbers
\usepackage{siunitx}

\title{High-flexibility reconstruction of small-scale motions in wall turbulence using a generalized zero-shot learning}

\author{Haokai Wu\aff{2},
  Kai Zhang\aff{1} \aff{2},
  Dai Zhou\aff{1} \aff{2},
  Wen-Li Chen\aff{3},
  Zhaolong Han\aff{1} \aff{2}
  \and Yong Cao\aff{1} \aff{2}
  \corresp{\email{yongcao@sjtu.edu.cn}}}

\affiliation{\aff{1}State Key Laboratory of Ocean Engineering, Shanghai Key Laboratory for Digital Maintenance of Buildings and Infrastructure, Shanghai Jiao Tong University, Shanghai, China
\aff{2}School of Ocean and Civil Engineering, Shanghai Jiao Tong University, Shanghai, China
\aff{3}School of Civil Engineering, Harbin Institute of Technology, Harbin, China}

\begin{document}
\maketitle
\begin{abstract}
This study proposes a novel super-resolution (or SR) framework for generating high-resolution turbulent boundary layer (TBL) flow from low-resolution inputs. The framework combines a super-resolution generative adversarial neural network (SRGAN) with down-sampling modules (DMs), integrating the residual of the continuity equation into the loss function. DMs selectively filter out components with excessive energy dissipation in low-resolution fields prior to the super-resolution process. The framework iteratively applies the SRGAN and DM procedure to fully capture the energy cascade of multi-scale flow structures, collectively termed the SRGAN-based energy cascade framework (EC-SRGAN). Despite being trained solely on turbulent channel flow data (via ``zero-shot transfer''), EC-SRGAN exhibits remarkable generalization in predicting TBL small-scale velocity fields, accurately reproducing wavenumber spectra compared to DNS results. Furthermore, a super-resolution core is trained at a specific super-resolution ratio. By leveraging this pre-trained super-resolution core, EC-SRGAN efficiently reconstructs TBL fields at multiple super-resolution ratios from various levels of low-resolution inputs, showcasing strong flexibility. By learning turbulent scale invariance, EC-SRGAN demonstrates robustness across different TBL datasets. These results underscore EC-SRGAN potential for generating and predicting wall turbulence with high flexibility, offering promising applications in addressing diverse TBL-related challenges.

\end{abstract}



\section{Introduction}
\label{sec:headings}

Turbulent boundary layers (TBL) play a crucial role in various engineering applications. Owing to the chaotic nature of turbulence and perturbations of wall friction, fully capturing the multi-scale TBL flow structures remains a challenging task.

The experimental measurements of TBL flow distribution are often insufficiently accurate to evaluate the small-scale flow characteristics in the inertial sub-range \citep{Shevkar23}. To fill the scale gap, direct numerical simulations (DNS) and large eddy simulations (LES) are introduced to reproduce the energy cascade of turbulence \citep{Cao22}. However, considerable computational costs are required to obtain physically realistic turbulence with sufficiently high mesh resolution. Therefore, in the field of turbulence prediction, a novel approach is highly motivated which focuses on both precision and efficiency. 
\begin{figure}
  \centering
  \includegraphics[width=0.9\linewidth]{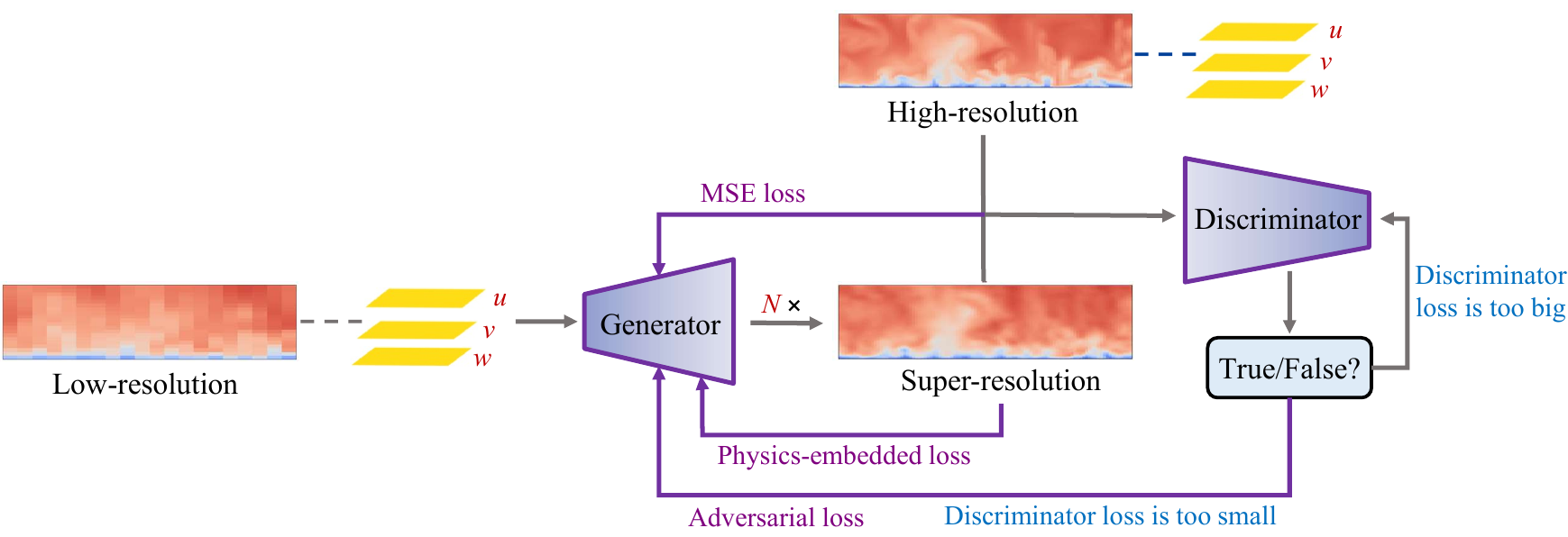}
  \caption{Schematic of physics-embedded SRGAN: overall training framework of SRGAN for neafr-wall turbulence reconstruction.}
  \label{fig:k1}
\end{figure}

With the availability of high-fidelity turbulence databases obtained from DNS, data-driven methods have rapidly advanced in deep learning algorithms (DLA), providing efficient solutions for turbulence reconstruction \citep{Vinuesa22}. From data with limited resolution alone, high-resolution (or HR) turbulence can be rapidly reconstructed using super-resolution DLA.

As a representative DLA, convolutional neural networks (CNNs) are commonly utilized to extract and integrate features from spatial flow distributions. \citet{Fukami21} efficiently reconstructed the wake flow field from extremely sparse measurements using CNN-based models. They also employed CNN convolution modules to construct super-resolution generative adversarial networks (SRGANs) for reconstructing turbulent fields. The effectiveness of the adversarial mechanism in SRGANs for enhancing the fidelity of reconstructed flow in the inertial range was demonstrated by \citet{Stengel20}. \citet{Güemes21} applied SRGANs to reconstruct near-wall turbulent velocity fields using coarse wall measurements. Compared with CNN-based method, SRGAN could provide a significant improvement in recovering the small-scale flow structures. \citet{Yousif21} trained the multi-scale enhanced SRGAN (MS-ESRGAN) with physics-based loss terms, including velocity gradients and Reynolds stress, to generate high-fidelity turbulent channel flow from sparsely distributed data. Their predictions of energy spectra showed compelling similarities with DNS results.

A transfer learning (one-shot learning) technique was employed to enhance the generalization capability of SRCNN or SRGAN-based models \citep{Guastoni21}. In one-shot learning, the model is fine tuned only with a limited amount of labeled data from the target (testing) dataset. Leveraging one-shot learning, the SRGAN-based model proposed by \citet{Yousif21} achieved acceptable accuracy in generating velocity fields across different Reynolds numbers. Similarly, \citet{Obiols21} utilized an SRCNN-based model called SURFNet to reconstruct super-resolution wakes around airfoils with diverse geometry boundaries.

However, current super-resolution efforts still have some notable limitations in practical fluid mechanics applications. Firstly, one-shot learning might fail due to insufficient training samples or limited data resolution in the turbulence data being predicted. If one-shot learning is not feasible, the question arises whether pre-trained models could be directly generalized to reconstruct turbulence in other related datasets. Secondly, most models can only reconstruct turbulent fields at a single super-resolution ratio. To achieve multiple super-resolution degrees of reconstruction, a bundle of models with different super-resolution ratios are required, leading to excessive consumption of training resources.

In response to the aforementioned challenges, this study proposes a framework that combines an SRGAN model (as the super-resolution core) with DMs. Its objective is to replicate the 2D super-resolution wall turbulence fields with a multi-scale energy cascade derived from low-resolution (or LR) ones. Thus, the proposed framework is referred to as the SRGAN-based energy cascade reconstruction framework (EC-SRGAN). EC-SRGAN is designed to flexibly adapt to inputs with various levels of coarse flow fields. Besides the high flexibility of inputs, ``zero-shot transfer'' \citep{Xian17} is employed instead of ``one-shot learning''. This approach involves training a model on a specific task using labeled data, but directly applying it to predict unseen datasets without any additional training or labeled data for the new task. It is particularly useful when acquiring labeled turbulence data for the new task is difficult or expensive.

\section{Methodology}\label{Methodology}
Focused on reconstructing multi-scale near-wall turbulence, the schematic explanations of the SRGAN model (super-resolution core of EC-SRGAN) and proposed EC-SRGAN are presented in Subsections 2.1 and Subsection 2.2, respectively.

\subsection{Physics-embedded SRGAN}
The inputs of the physics-embedded SRGAN model are flow velocity fields with low-resolution grids of $m/N$ × $n/N$ in the horizontal and vertical directions, respectively. Its objective is to accurately output the $m$ × $n$ super-resolution near-wall turbulence distributions. Hereafter, the $N$ × $N$ times finer super-resolution model is denoted as $N$× SRGAN. The model plays the core role of increasing the flow field dimensions within a EC-SRGAN framework.
 
This study rebuilds the SRGAN following the architecture proposed by \citet{Wang19} that was originally dedicated to image super-resolution. The inner structures of generator (G) and discriminator (D) have been extensively elucidated by \citet{Wang19} and \citet{Wu23}.

As shown in figure \ref{fig:k1}, the training loss function of G is composed of four contributions. The first two contributions are mean squared error ($L_{MSE}$) and adversarial loss ($L_{Adver.}$). The $L_{MSE}$ is computed based on the pixel difference between the true (high-resolution label) and reconstructed (super-resolution output) turbulent fields:
\begin{equation}
  L_{MSE}=\left \| y-G(x) \right \| _{2}^{2},
\end{equation}
The second part apply $L_{Adver.}$ of G as follows:
\begin{equation}
  L_{Adver.}=-log(D(G(x)),
\end{equation}
where $D(G(x))$ outputs the probability that the super-resolution field originates from the high-resolution data \citep{Stengel20}. During the training, the weights in G are updated in the direction that $D(G(x))$ returns a probability value approaching to 1, so that the reconstructed flow distributions could be similar to the true ones. This loss term could measure the generator’s capability of “fooling” the discriminator, with a lower value indicating more robust performance of G.

To reconstruct physically realistic turbulence and better identify the sharp gradients, this study embeds a priori information into the training process. That is, the residual of the continuity equation of 3D incompressible flow and the errors of velocity gradients are selected as the third and fourth contributions of the loss function, respectively:
\begin{equation}
  L_{Conti.}=(\partial v_i/\partial x_i)_{SR}, (i=1, 2, 3),
\end{equation}
\begin{equation}
  L_{Grad.}= \textstyle \sum_{i=1}^{3} \left |(\partial v_i/\partial x_i)_{HR}-(\partial v_i/\partial x_i)_{SR}\right |,
\end{equation}
where the derivatives $\partial v_i/\partial x_i$ ($\partial u/\partial x$, $\partial v/\partial y$ and $\partial w/\partial z$) are calculated by finite difference method and all discretized automatically during the training. The subscripts ``HR'' and ``SR'' denote that the velocity gradients are calculated from the reference and reconstructed flow fields, respectively. Since the physical quantities are all nondimensionalized from the datasets we utilize, it is reasonable to combine the loss terms with different meanings into the loss function. The total loss function $L_G$ of G is hence defined as:
\begin{equation}
  L_G=L_{MSE}+\alpha \cdot  L_{Adver.}+\beta_1 \cdot  L_{Conti.}+\beta_2 \cdot  L_{Grad.},
\end{equation}
where $\alpha$, $\beta_1$ and $\beta_2$ are balance factors to scale magnitudes of respective loss terms. The $\alpha$ is fixed as $10^{-3}$ in accordance with the application of \citet{Stengel20}. To fully integrate a priori information into the trained model, $\beta_1$ and $\beta_2$ should be adjusted dynamically such that the combined influence of two physics-embedded terms constitutes more than $50\%$ of $L_G$ at each training epoch. The proportion $50\%$ is empirically selected to ensure that the weights of physics-related loss contributions are not less than those of $L_{MSE}$ and $L_{Adver.}$. This aims to learn the conservation law and gradient characteristics fully for the SRGAN model \citep{Wu23}.

During the training process, the ADAM optimizer is utilized to minimize the total loss of G with a learning rate of $10^{-4}$. In the meantime, it is imperative to monitor the discriminator loss of D ($L_D$), according to the training procedure in \citet{Stengel20}. The training process described above is implemented based on Tensorflow-1f.15-GPU.

\subsection{EC-SRGAN framework}
\begin{figure}
  \centering
  \includegraphics[width=1.0\linewidth]{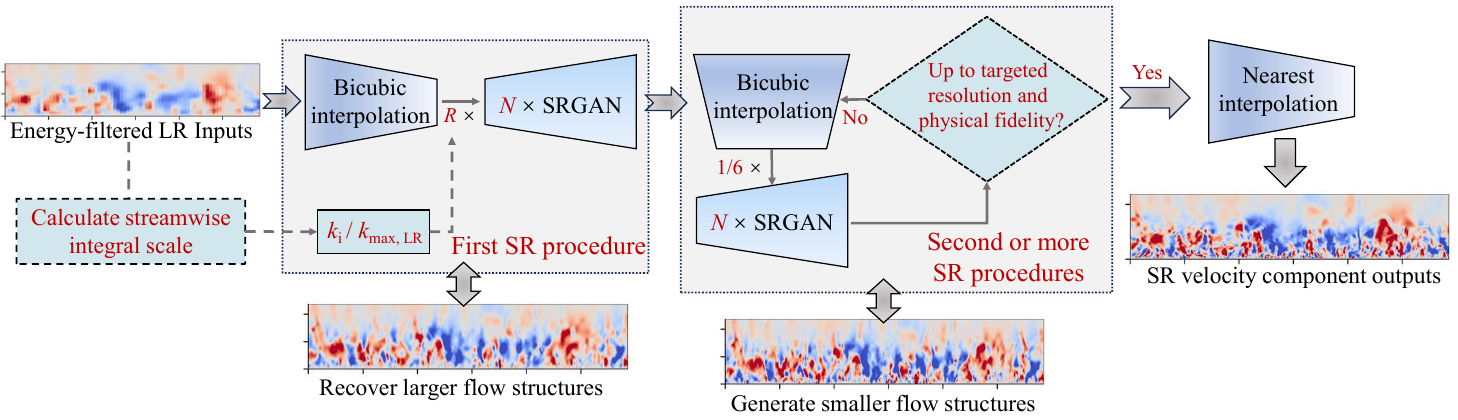}
  \caption{Schematic workflow of the EC-SRGAN framework for multi-scale reconstruction of TBL. Two or more super-resolution (or SR) procedures are included in the framework.}
  \label{fig:k2}
\end{figure}

Although the SRGAN model increases the data dimensionality of the flow velocity, it cannot guarantee the reproduction of multi-scale structures in the super-resolution flow fields. To realize multi-scale prediction, a reasonable framework design is further needed based on the trained SRGAN model.

Figure \ref{fig:k2} shows the schematic workflow of the EC-SRGAN framework employed in reconstructing multi-scale TBL. As the inputs of EC-SRGAN, the low-resolution fields of velocity components are initially obtained by average-pooling from the DNS TBL fields. Next, low-pass filtering is applied to the initial low-resolution fields. This imitates the energy dissipation of small-scale flow structures when the TBL simulation is conducted in the low-resolution grids. After the above down-sampling processes, the energy-containing region should be encompassed by the wavenumber spectra of low-resolution input data. This ensures that the averaged scale characteristics of flow fields, including turbulent integral scale, could be appropriately restored and captured. 

To successively recover the turbulence energy of wall turbulence from large to small scales, the filtered low-resolution inputs are subjected to two or more super-resolution procedures. In each procedure, the trained $N$× SRGAN model serves as the core for up-sampling the turbulence. Prior to the up-sampling, down-sampling modules are incorporated to filter out flow components exhibiting excessive energy dissipation in the low-resolution inputs. This step presents an opportunity for the super-resolution procedure to accurately reproduce flow structures at larger scales. Once the energy recovery is realized at larger scales, it is promising that the SRGAN model would progressively infer the smaller-scale flow structures, leveraging the energy cascade theory.

In the first super-resolution procedure, the streamwise integral scale ($l_x$) of the low-resolution velocity fields is first calculated. Next, the bicubic interpolation module is used to further down-sample the low-resolution fields to sparser grid distributions, where the maximum wavenumber ($k_{xm}$) nearly corresponds to the $l_x$. Through this initial down-sampling interpolation, flow scales smaller than $l_x$ are filtered out from the low-resolution inputs. Following this step, the trained $N$× SRGAN model is utilized to expand the dimensions of the flow fields. The goal of this procedure is to initially reproduce the absent flow structures at larger scales within energy-containing and early inertial subregions. 

In the second super-resolution procedure, the initially reconstructed fields are further down-sampled by another bicubic interpolation, with a ratio of 1/6. The ratio is determined based on the approximation that the Taylor scale is typically around $1/6$ of the integral scale for high $\Rey$ turbulence. Then, the down-sampled fields are reconstructed once more by the trained up-sampling core. It is expected to further recover turbulence energy at larger scales and generate spatial fluctuations at smaller scales within inertial subregion. Through this procedure, the grid dimension of the flow velocity data is expanded by (1/6×$N$) times. 

As for multi-ratio super-resolution reconstruction from different levels of low-resolution inputs, the secret lies in adjusting both the number of super-resolution procedures and the down-sampling ratio of final nearest interpolation. The (1/6×$N$)× super-resolution procedure should be reconducted, until following two conditions are both satisfied. Firstly, the reconstructed flow fields exceed the targeted resolution of high-resolution ones. Secondly, for the super-resolution outputs, the cut-off wavenumber of inertial subregion is larger than 6 times the $k_{xm}$ corresponding to $l_x$. As a final process, super-resolution outputs are required to align with the targeted resolution. Thus, the final down-sampling ratio could be determined. The reconstruction fidelity of super-resolution flow fields is verified through the comparison with high-resolution ones from the DNS dataset.

\section{Data description and processing}\label{Data}
In terms of the turbulence generation mechanism, channel flow and boundary layer exhibit similarities as they are both influenced by wall friction. Therefore, this study plans to apply turbulent channel flow database \citep{Lee13} available at the Johns Hopkins turbulence databases (JHTDB) to exclusively train and validate the SRGAN model. Furthermore, ``zero-shot transfer'' performance of the trained SRGAN and its carrier (EC-SRGAN) is evaluated based on TBL database \citep{Lee18} from JHTDB.

The selected turbulent channel flow dataset was obtained from DNS. The coordinates $x$, $y$ and $z$ are defined as the streamwise, wall-normal and crossflow, respectively, with the corresponding velocity components $u$, $v$ and $w$. Flow fields of these components are stored at a dimensionless time step of 0.0065 in a duration of $t$=[0, 25.9935], resulting in 4000 snapshots of data. 

Focused on the reconstruction of near-wall turbulence, training and validating data are thus obtained from the 2D planes normal to the streamwise direction ($y$–$z$ planes). The domain size of training and validating datasets is fixed to be $y$ × $z$ = 0.3835$H$ × 1.5340$H$ with 500 × 500 uniformly distributed grid nodes extracted, where $H$ is the channel height. The grid is dense enough to resolve the small-scale characteristics of turbulent channel flow. $y$–$z$ planes of $x = \pi H$ and $x = 3\pi H$ are chosen as the sources of training and validating datasets, respectively. In the training process, to calculate $\partial u/\partial x$ in the $L_{Conti.}$, velocity components $u$ in the neighboring $y-z$ planes of $x = \pi H$ and $x = 3\pi H$ are both required. Thus, the data from $x = \pi H + \delta x$ and $x = 3\pi H + \delta x$ are respectively added to the training and validating datasets, where $\delta x$ is the streamwise grid spacing used in DNS. However, since the $\partial v/\partial y$ and $\partial u/\partial x$ could be discretized within one $y-z$ plane, there is no need to extract extra $v$ and $w$ data from two neighboring planes. 

At each instance, it is necessary to establish a mapping between high-resolution and low-resolution velocity component data for SRGAN model. As the output labels of the SRGAN model, high-resolution fields are organized into the dimension of 500 × 500 × 4, where ``500 × 500'' is the grid size of high-resolution field, and ``4'' indicates the channel number of the output. The first two channels store the distributions of velocity component $u$ from two neighboring planes, while $v$ and $w$ fields are placed in the last two channels, respectively. Fed as inputs of $N$ × SRGAN model, the low-resolution velocity components are obtained by average pooling from corresponding channels of high-resolution labels to a dimension of 500/$N$ × 500/$N$ × 4. Hereafter, the low-resolution input is termed as 1/$N$× input.

Alternatively, the TBL dataset provided by JHTDB is employed as the testing set. The dataset was produced via DNS of incompressible flow over a no-slip flat plate. Note that the definitions of the velocity components are the same as those used in turbulent channel flow dataset. The detailed descriptions on domain and grid sizes were presented in \citet{Lee18} and \citet{Yousif23}. The available data span a period of $t=[0, 1175]L/U$ with a time step of $0.25L/U$ where $L$ represents the half-thickness. Thus, the dataset here contains 4700 snapshots for EC-SRGAN testing.

The testing set is constituted by the velocity fields from only one targeted $y$–$z$ plane of $x = 1000L$, where the turbulence was verified to be fully developed \citep{Lee18}. The size of the testing 2D zone is set as $y$ × $z$ = $23.65L$ × $117.2L$, corresponding to 200 × 400 grid nodes extracted at an equal spacing in each direction. As a high-resolution label, the data dimension is 200 × 400 × 4 for each testing sample. The size of the high-resolution data is further reduced by 1/$N$ to 200/$N$ × 400/$N$ × 4, which could be realized by average pooling and low-pass filtering, as depicted in Subsection 2.2.

\section{Results and discussion}\label{Results}
\subsection{Validations of turbulent channel flow reconstruction based on physics-embedded SRGAN}

This subsection validates the capability of a 12× physics-embedded SRGAN model to reconstruct the super-resolution channel flow fields from 1/12× low-resolution data of velocity components. This model would be inserted into the EC-SRGAN as a super-resolution core in last three subsections. 

\begin{figure}
  \centering
  \includegraphics[width=1.0\linewidth]{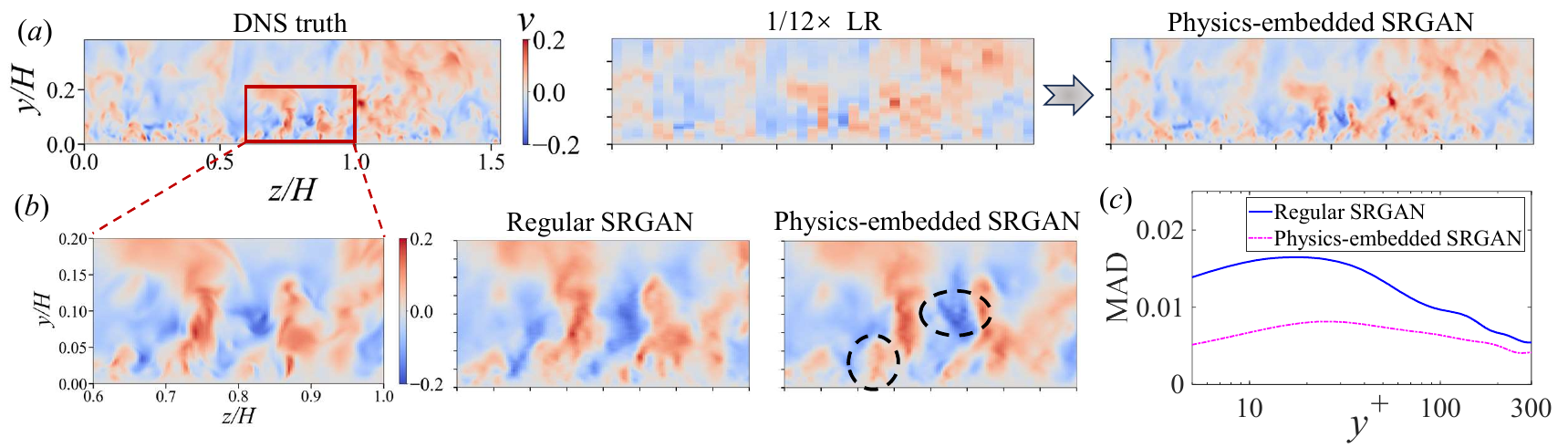}
  \caption{Reconstruction performances of the physics-embedded SRGAN models: instantaneous turbulent channel flow fields of components $v$ over (a) validating plane and (b) the extracted subdomain when $t = 13.0$ and (c) wall-normal variations of mean absolute divergences of super-resolution fields within the validating period.}
  \label{fig:k4}
\end{figure}

In figure \ref{fig:k4}(a), the instantaneous super-resolution velocity field of components $v$ is compared with the DNS snapshots at the same instant of $t=13.0$. The physics-embedded SRGAN model could successfully generate the velocity fluctuations at small scales that are seriously missing in the low-resolution input near the wall. The effect of the embedded physical loss terms on SRGAN training process is further assessed according to the velocity distributions (figure \ref{fig:k4}(b)). A subdomain is extracted near the wall for detailed comparisons. With the help of embedded $L_{Conti.}$ and $L_{Grad.}$, the rolling patterns of flow structures are reproduced better compared with the result from the regular model, as highlighted by the black ellipses in figure \ref{fig:k4}(b).

Two indicators are utilized to assess the reconstruction performances of regular and physics-embedded SRGAN models. As shown in table \ref{tab:k0}, the mean square error (MSE) and determination coefficient ($R^2$) both represent the agreement degree of spatial fluctuations between the reconstructed and DNS total velocity ($\sqrt{u^2+v^2+w^2}$) fields. As physical loss terms are informed, the physics-embedded SRGAN model achieves lower MSE and higher $R^2$ indicating even stronger fitting capacity for complex velocity fluctuations. To further reflect the impact of physical loss terms, the mean absolute divergence (MAD) reflecting the residuals of the continuity equation is assessed in figure \ref{fig:k4}(c). The MAD is obtained by two steps. Firstly, it is computed on each grid node in the reconstructed flow fields averaged over 4000 time steps (given by the formula of $1/4000\sum \left | \nabla \cdot \boldsymbol{v}\right |$). Secondly, the time-averaged divergences are further averaged across the nodes at the same height away from the wall surface. Smaller MAD values approaching to 0 indicate a closer match between the reconstructed and true incompressible flow fields. As shown in figure \ref{fig:k4}(c), below the log-law region ($y^+<30$), the divergences computed from the regular SRGAN are 2 to 3 times as large as those from the physics-embedded model in the corresponding height. The current physics-embedded approach is proved to improve the physics-based fidelity of reconstructed velocity gradient distributions for the SRGAN model.

\subsection{``Zero-shot'' reconstruction of TBL only based on 5× super-resolution core}

\begin{table}
  \begin{center}
\def~{\hphantom{0}}
  \begin{tabular}{lcc}
      Model             & MSE             & $R^2$   \\[3pt]
      Regular           & $2.25$×$10^{-1}$        & 0.9715 \\
      Physics-embedded  & $8.96$×$10^{-2}$        & 0.9864 \\
  \end{tabular}
  \caption{Comparison of reconstruction performances of regular and physics-embedded SRGAN models.}
  \label{tab:k0}
  \end{center}
\end{table}

\begin{figure}
  \centering
  \includegraphics[width=0.9\linewidth]{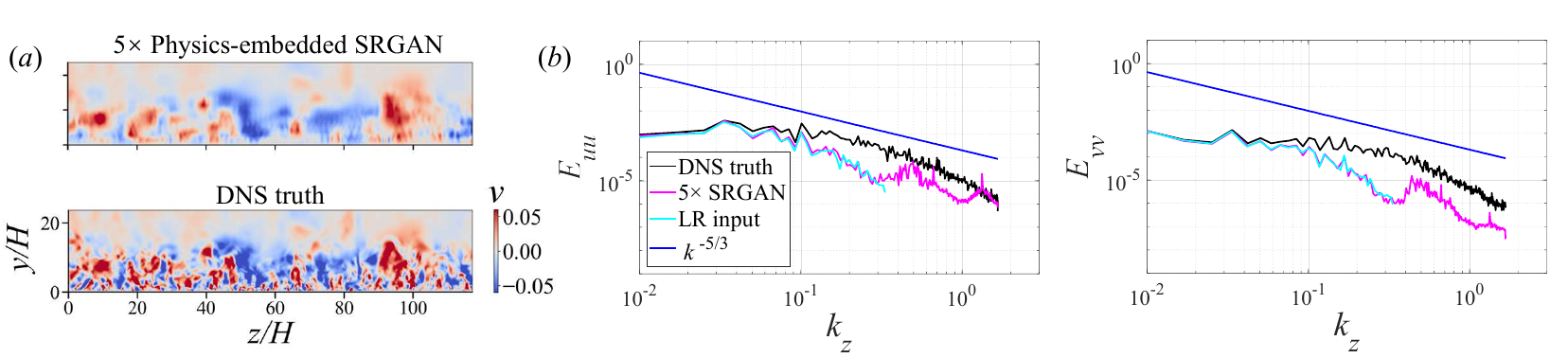}
  \caption{``Zero-shot" transfer performances of the physics-embedded SRGAN model in TBL reconstruction: (a) instantaneous super-resolution velocity fields of component $v$ when $t = 727.5$ and (b) wavenumber spectra of the instantaneous $u$ and $v$ fields, compared with the corresponding results of high-resolution and low-resolution (or LR) fields.} 
  \label{fig:k5}
\end{figure}

\begin{figure}
  \centering
  \includegraphics[width=1.0\linewidth]{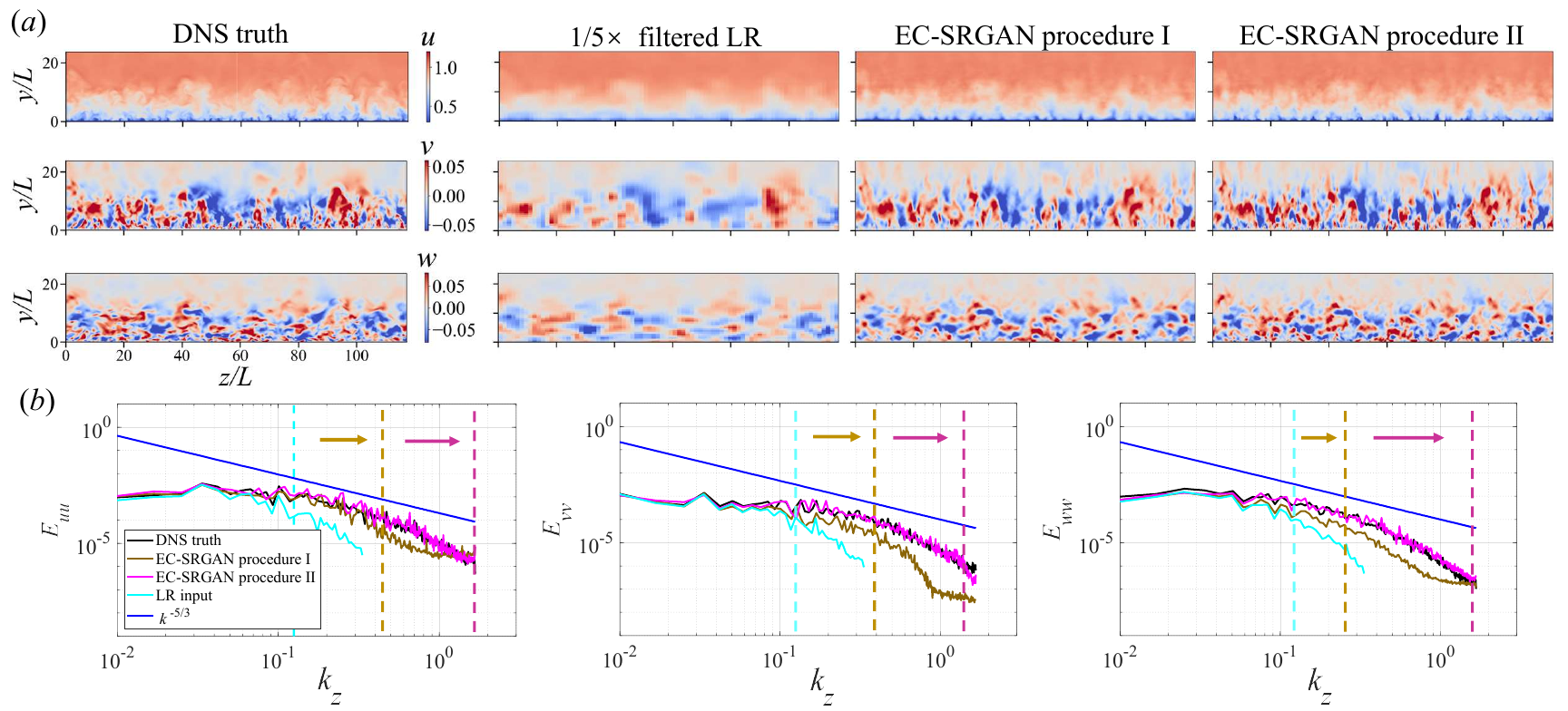}
  \caption{``Zero-shot" transfer performances of EC-SRGAN framework in (a) instantaneous TBL reconstructions for three velocity components when $t = 727.5$ and (b) wavenumber spectra of the instantaneous fields through two SR procedures.}
  \label{fig:k6_1}
\end{figure}

To evaluate the applicability of the proposed ``zero-shot'' approach, the TBL fields are intended to be reconstructed over the plane from the energy-filtered low-resolution inputs. The filtered inputs could be treated as the numerical simulation results from the low-resolution mesh. Without any transfer training, the physics-embedded SRGAN model, i.e., super-resolution core in EC-SRGAN, is directly utilized to predict the super-resolution velocity fields of TBL (``zero-shot'' transfer).

Here, the 5× super-resolution of TBL is taken as the first testing case for model transfer. Figure. \ref{fig:k5}(a) illustrates the super-resolution wall-normal velocity ($v$), along with the low-resolution and DNS fields at the same instant ($t = 727.5$). Disappointingly, the super-resolution field exhibits severe magnitude underestimations and non-physical fluctuations. Figure. \ref{fig:k5}(b) presents wavenumber spectra of the $u$ and $v$ fields at $t = 727.5$ as function of $k_z$, which elucidate horizontal distributions and energy cascade characteristics of turbulent structures near the wall. In lower-wavenumber region, the spectra of super-resolution $u$ and $v$ fields maintain similar dissipation trends with those of the corresponding low-resolution inputs, violating the -5/3 power law in the inertial subregion. They both rebound around the cut-off wavenumbers of low-resolution inputs and maintain higher energy levels within the higher-wavenumber region, accounting for the odd fluctuations occurred in figure \ref{fig:k5}(a). The predicted spectra of streamwise and non-streamwise velocities jointly reflect that ``zero-shot'' transfer relying solely on a super-resolution core cannot generate the physically realistic turbulence conforming to the energy cascade theory.

\begin{figure}
  \centering
  \includegraphics[width=1.0\linewidth]{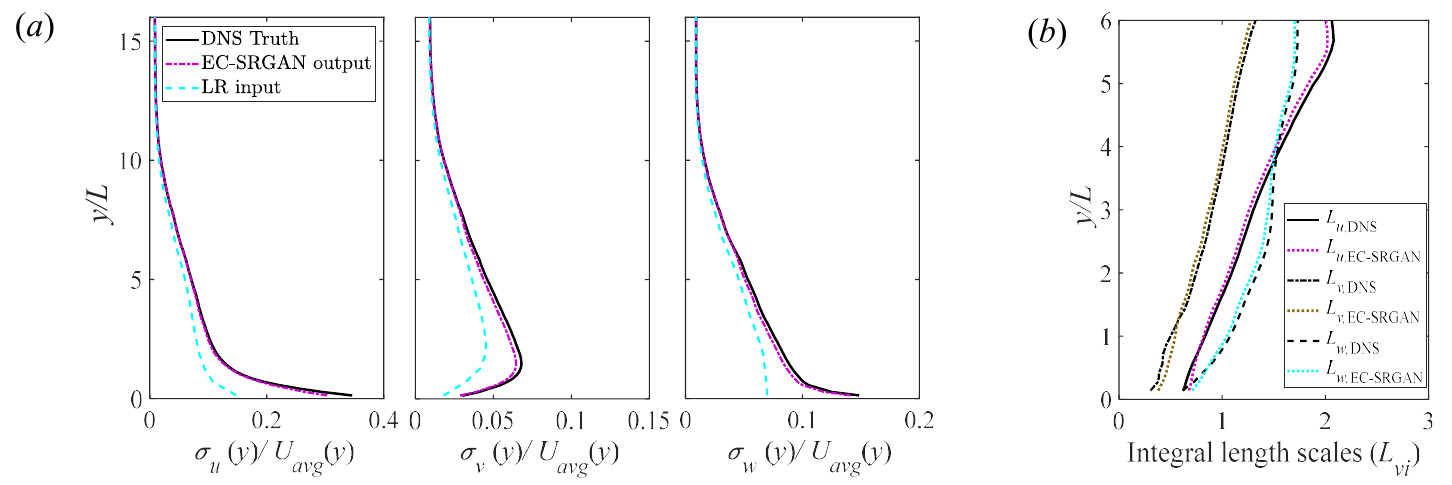}
  \caption{``Zero-shot" transfer performances of EC-SRGAN framework in TBL reconstructions for vertical profiles of (a) turbulent intensity and (b) horizontal integral length scales of three velocity components.}
  \label{fig:k6_2}
\end{figure}

\begin{figure}
  \centering
  \includegraphics[width=1.0\linewidth]{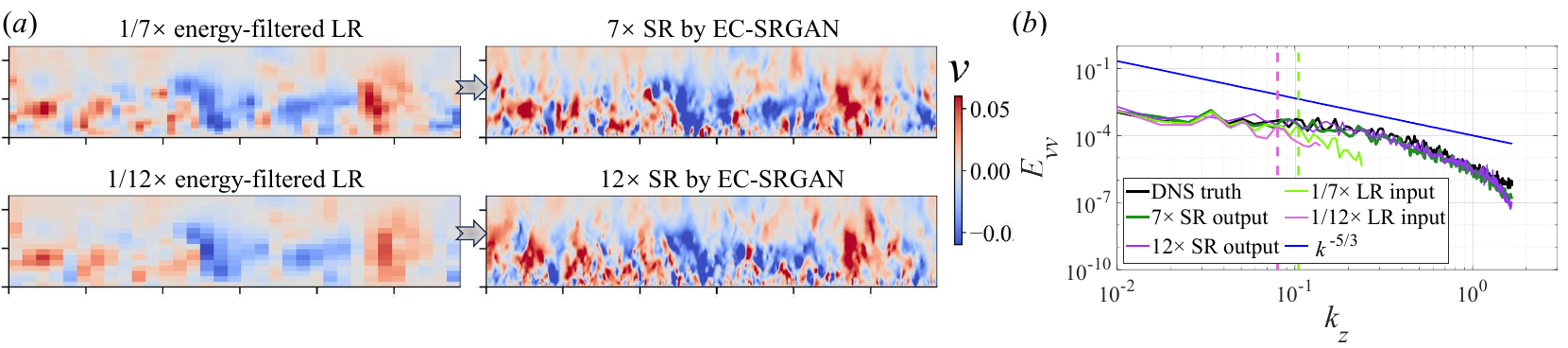}
  \caption{Flexibility examination of multi-ratio reconstruction by EC-SRGAN: (a) reconstructed instantaneous $v$ fields from 1/7× and 1/12× low-resolution (or LR) inputs and (b) wavenumber spectra of corresponding super-resolution (or SR) outputs compared with the results of DNS field.}
  \label{fig:k7}
\end{figure}

\subsection{``Zero-shot'' reconstruction of TBL based on 5× EC-SRGAN}

The subsection delves into the applicability of ``zero-shot'' reconstruction using EC-SRGAN. An super-resolution core is inserted into EC-SRGAN to enforce the 5× reconstruction. Within EC-SRGAN, the down-sampling level of $R$× is set as 5/12 in the initail bicubic interpolation module, following the principle introduced in Subsection 2.2. Subsequently, the first super-resolution procedure involves feeding the down-sampled low-resolution fields into the SRGAN model. Instead of the 5× SRGAN model, the trained 12× model is informed in the EC-SRGAN, anticipating reproducing more multi-scale structures from the low-resolution inputs. Figure. \ref{fig:k6_1}(a) shows the instantaneous TBL of three velocity components reconstructed by EC-SRGAN. Through the first super-resolution procedure, the reconstructed patterns of large-scale velocity structures match well with those of DNS fields. 

Figure. \ref{fig:k6_1}(b) exihibits the horizontal wavenumber spectra of three velocity components. Notably, the turbulent energy recovers towards that of DNS results during the first reconstruction. The -5/3 downward trends persist up to 2-4 times the wavenumber corresponding to $l_x$ of the low-resolution inputs ($\thicksim1.2$×$10^{-1}$), after which dissipations occur as wavenumbers increase. It suggests that the first super-resolution procedure primarily focuses on initially reproducing velocity structures and recovering turbulent energy at larger spatial scales, i.e., the energy-containing and early inertial subregion.

The second procedure is hence required for enhancing the recovery of turbulent energy and replicating the smaller-scale velocity fluctuations. This procedure achieves a (1/6×12)× super-resolution, followed by a 1/2 nearest interpolation to align the resolutions. Firstly, the instantaneous flow fields are qualitatively compared. In figure \ref{fig:k6_1}, the second super-resolution procedure notably enhances the generation of smaller-scale structures near the wall, particularly for the wall-normal and crossflow velocity components. 

Then, the turbulent intensities of reconstructed fields are calculated. Figure. \ref{fig:k6_2}(a) respectively presents the vertical profiles of streamwise, wall-normal and transverse turbulence intensity ($\sigma_u(y)/U_{avg}(y)$, $\sigma_v(y)/U_{avg}(y)$ and $\sigma_w(y)/U_{avg}(y)$). During the testing period, the turbulence statistics are calculated over 4000 time-steps to reach the stationary states. Through two super-resolution procedures, as indicated in figure \ref{fig:k6_2}(a), three reproduced profiles of EC-SRGAN are all generally consistent with spatial distributions of turbulence intensities in the DNS data. From 1/5× low-resolution inputs, the EC-SRGAN could significantly recover the high turbulent intensity characteristics in the near-wall region approaching to those of the DNS fields, although there are slight magnitude deficits. The recovery of high turbulent intensities facilitates the EC-SRGAN recognizing the kinetic energy and momentum transport of near-wall vortex structures, which are intricately linked to the turbulent scale characteristics. Therefore, the multiple turbulent length scales are expected to reasonably inferred from the accurate reproductions of turbulence intensities.

Next, to check the reconstruction on the averaged turbulent scales, the verical profiles of horizontal integral length scales are compared between the DNS and reconstructed flow fields in figure \ref{fig:k6_2}(b). As shown in figure \ref{fig:k6_2}(b), three length scales in the reconstructed fields all match well with those in DNS references, through two super-resolution procedures of EC-SRGAN. It indicates that EC-SRGAN achieves ideal prediction accuracy on average size of the most energetic turbulent eddies.

Finally, we analyze the small-scale reconstruction of the turbulent energy. In figure \ref{fig:k6_1}(b), through two super-resolution procedures, the turbulent energy demonstrates continuous recovery at larger scales, closely matching the DNS references. Furthermore, at smaller scales, the predicted end of the inertial subrange extends beyond 2-4 times to over 10 times the wavenumber corresponding to the $l_x$. This expansion signifies that, through a minimum of two reconstruction steps, the small-scale velocity characteristics could be physically preserved from the high-fidelity TBL fields.

\begin {comment}
\begin{table}
  \begin{center}
\def~{\hphantom{0}}
  \begin{tabular}{lcc}
  Velocity components & DNS references  & EC-SRGAN outputs \\[3pt]
        Component $u$           & 2.0505        & 1.9783 \\
        Component $v$           & 1.7762        & 1.7266 \\
        Component $w$           & 2.3012        & 2.1848 \\
  \end{tabular}
  \caption{Comparison of averaged horizontal integral length scales of three velocity components between DNS and reconstructed flow fields.}
  \label{tab:k2}
  \end{center}
\end{table}
\end {comment}

\subsection{Flexibility examination of multi-ratio reconstruction by EC-SRGAN}

Furthermore, to extend the flexibility of EC-SRGAN, it is expected that the EC-SRGAN could realize the TBL reconstruction with multiple super-resolution ratios from different levels of coarse turbulent fields. Figure. \ref{fig:k7}(a) and (b) illustrate instantaneous $v$ fields reconstructed by EC-SRGAN from 1/7 and 1/12× low-resolution inputs. Superior to other SRGAN-based reconstruction tests of wall turbulence \citep{Güemes21}, the EC-SRGAN could maintain similar reconstruction performances, even though it is input by low-resolution fields with different filtering levels.

Figure. \ref{fig:k7}(c) presents the wavenumber spectra of the wall-normal velocity fields reconstructed from 1/7 and 1/12× low-resolution inputs by the EC-SRGAN. Despite slight energy loss, the framework succeeds in replicating the -5/3 law of energy cascade in the inertial subrange, irrespective of the low-resolution field levels provided. It can be concluded that EC-SRGAN can accommodate multiple super-resolution tasks to reconstruct multi-scale structures with high physical fidelity. Moreover, the proposed super-resolution framework exhibits commendable robustness in TBL reconstruction against the mesh coarseness of low-resolution fields.

\begin{table}
  \begin{center}
\def~{\hphantom{0}}
  \begin{tabular}{lcccc}
      Resources  & Method   & $Re_{\theta}$ of selected plane & $R^2$ & MAD   \\[3pt]
       JHTDB   & DNS & $\thicksim1300$ & 0.9543 & $5.65$×$10^{-3}$ \\
       \citet{Towne23}   & DNS & $\thicksim1900$ & 0.9216 & $5.47$×$10^{-3}$ \\
  \end{tabular}
  \caption{Comparison of reconstruction performances of EC-SRGAN between different TBL datasets.}
  \label{tab:k1}
  \end{center}
\end{table}

\begin{figure}
  \centering
  \includegraphics[width=1.0\linewidth]{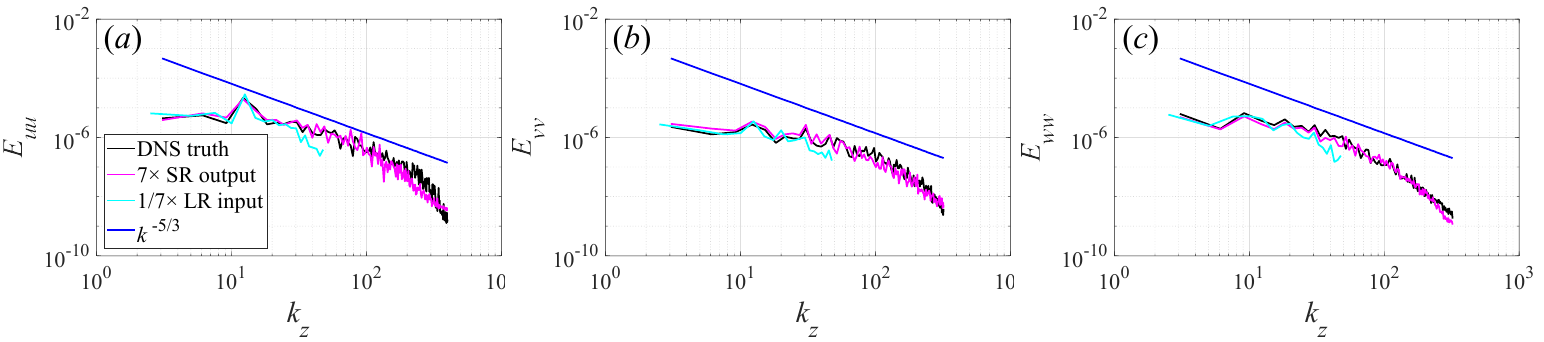}
  \caption{Wavenumber spectra of the reconstructed (or SR) instantaneous fields: (a) $u$, (b) $v$ and (c) $w$ in TBL dataset of \citet{Towne23}, compared with the corresponding low-resolution (or LR) inputs and high-resolution results.}
  \label{fig:k8}
\end{figure}

\subsection{Generalization capability of EC-SRGAN among different datasets}

Finally, to verify the generalization capability of EC-SRGAN, reconstruction performances between different TBL datasets are compared in Table \ref{tab:k1}. Besides data from JHTDB ($Re_{\theta}\thicksim1300$, based on momentum thickness $\theta$), another TBL plane is selected from a DNS dataset \citep{Towne23} for TBL reconstruction testing, which has even higher $Re_{\theta}$ ($\thicksim1900$). Reconstruction performances are evaluated by $R^2$ and MAD, where the MAD computed across shear dominating region ($y^{+}<300$) reflects the reproducibility of the incompressible flow. As shown in table \ref{tab:k1}, for an alternative dataset with higher $Re_{\theta}$, the current reconstruction framework can achieve good fitting of velocity fluctuations ($R^2>0.9)$ and maintain similar physical fidelity (similar MAD). Figure. \ref{fig:k8} exhibits the wavenumber spectra of three velocity component fields reconstructed from 1/7× low-resolution inputs by the EC-SRGAN. For velocity data at higher $Re_{\theta}$, the energy cascade in the inertial subregions can still be well reproduced, although the energy in the dissipation region is slightly underestimated. 

We further analyzed the reason why the current method shows remarkable generalization capability across various TBL datasets. Initially, it is essential to acknowledge that turbulence contains similarities in vortex structures across a spectrum of spatial scales. Fortunately, in the SRGAN model within EC-SRGAN, the convolutional filters burden the task to compress and expand the multi-scale distributions of turbulence fields. They play a pivotal role in nonlinearly capturing the data features linked to the invariants of the velocity gradient tensor $Q$ and $R$. Once the nonlinear invariants are identified, the super-resolution procedures can effectively reconstruct flow structures with sufficient strain and vortex stretching. Furthermore, through the multi-scale feature extraction, the energy cascade in TBL could be accurately represented by the super-resolution procedures \citep{Mi23}. This, in turn, contributes to enhancing the similarity of spatial distributions across different Reynolds numbers. Analogously, the scale-invariant features are also incorporated during the image feature extraction using CNN \citep{Fukami24}. Therefore, across various TBL datasets, the similarities of multi-scale vortical structures could be well-preserved, showcasing high robustness of the EC-SRGAN.

\section{Conclusions}\label{Conclusions}
This study proposes EC-SRGAN, a flexible framework designed to reconstruct small-scale motions in wall turbulence through zero-shot transfer. The framework employs iterative super-resolution procedures to reconstruct multi-scale turbulent structures. The model is trained once using the turbulent channel flow database and then applied to predict the TBL field. Firstly, EC-SRGAN demonstrates remarkable transfer capability by accurately predicting the instantaneous velocity fields of TBL and reproducing wavenumber spectra with high fidelity compared to DNS results. Secondly, the framework exhibits high flexibility in reconstruction from various levels of coarse flow fields while preserving physical accuracy. Lastly, EC-SRGAN possesses high robustness in TBL reconstruction across different TBL datasets, likely due to its incorporation of scale invariance in turbulence.

\backsection[Acknowledgements]
{This study was funded by the National Key R\&D Program of China (2023YFE0120000), the National Natural Science Foundation of China (52108462, 52122110), the Shanghai Sailing Program (21YF1419400), the Natural Science Foundation of Shanghai (21ZR1428900), the Natural Science Foundation of Chongqing (CSTB2023NSCQ-MSX0060) and supported by the ORISE Supercomputer.}

\backsection[Declaration of interests]
{The authors report no conflict of interest.}

\bibliographystyle{jfm}


\end{document}